\documentclass{book}    
\usepackage{piers}  
\usepackage{amsmath}

\usepackage{color}

\def\>{{\rangle}}
\def\<{{\langle}}
\def\)>{{)\!\rangle}}
\def\(<{{\langle\!(}}

\begin{document}

\title{Analysis of high-harmonic generation in terms of complex Floquet spectral analysis}
\maketitle

\author      {H. Yamane}
\affiliation {Osaka Prefecture University}
\address     {}
\city        {Osaka}
\postalcode  {}
\country     {Japan}
\phone       {}    
\fax         {}    
\email       {}  
\misc        { }  
\nomakeauthor

\author      {S. Tanaka}
\affiliation {Osaka Prefecture University}
\address     {}
\city        {Osaka}
\postalcode  {}
\country     {Japan}
\phone       {}    
\fax         {}    
\email       {}  
\misc        { }  
\nomakeauthor
\author      {M. Domina}
\affiliation {University of Palermo}
\address     {}
\city        {Palermo}
\postalcode  {}
\country     {Italy}
\phone       {}    
\fax         {}    
\email       {}  
\misc        { }  
\nomakeauthor
\author      {R. Passante}
\affiliation {University of Palermo}
\address     {}
\city        {Palermo}
\postalcode  {}
\country     {Italy}
\phone       {}    
\fax         {}    
\email       {}  
\misc        { }  
\nomakeauthor
\author      {T. Petrosky}
\affiliation {Osaka Prefecture University}
\address     {University of Texas at Austin}
\city        {Austin}
\postalcode  {}
\country     {USA}
\phone       {}    
\fax         {}    
\email       {}  
\misc        { }  
\nomakeauthor

\begin{authors}

{\bf H. Yamane}$^{1}$, {\bf S. Tanaka}$^{1}$,  {\bf  M. Domina}$^{2}$, {\bf  R. Passante}$^{2}$, {\bf and T. Petrosky}$^{3,4}$\\
\medskip
$^{1}$Department of Physical Science, Osaka Prefecture University, Japan\\
$^{2}$Department of Physics and Chemistry, University of Palermo, Italy\\
$^{3}$Center for Complex Quantum System, University of Texas at Austin, USA\\
$^{4}$Institute of Industrial Science, The University of Tokyo, Japan

\end{authors}

\begin{paper}

\begin{piersabstract}
Recent developments on intense laser sources is opening a new field of optical sciences.
An intense coherent light beam strongly interacting with the matter causes a coherent motion of a particle, forming a strongly dressed excited particle. 
A photon emission from this dressed excited particle is a strong nonlinear process causing high-harmonic generation(HHG), where the perturbation analysis is broken down. 

In this work, we study a coherent photon emission from a strongly dressed excited atom in terms of complex spectral analysis in the extended Floquet-Hilbert-space.
We have obtained the eigenstates of the total Hamiltonian with use of Feshbach-Brilloiun-Wigner projection method.
In this extended  space, the eigenstates of the total Hamiltonian consisting of the radiation field and the atom system have complex eigenvalues whose imaginary part represents the decay rate.
Time evolution of the system is represented by the complex eigenvector expansion so that the correlation dynamics between the photon and the atom is fully taken into account.  
The HHG is interpreted as the irreversible spontaneous photon emission due to the resonance singularity in terms of the multiple Floquet states that are generated by periodic external field.
We have found that the interference between the emitted photons over the different Floquet states causes spatial pulse emission correlated with the decay process of the excited atom.

\end{piersabstract}

\psection{Introduction}

Owing to the development of intense laser light sources, there has been made a great progress in the quantum control of atoms and molecules.\cite{Silveri17RPP,Shapiro06PhysRep,Kohler05PysRep,Grifoni98PhysRep}
An induced coherent motion of materials then generates the high-harmonic radiation whose wavelength has reached to x-ray region.\cite{Krause92PRL,Faisal97PRA,Ghimire11NatPhys,Joachain11Book}

It has been intriguing to investigate the microscopic mechanism of high-harmonic generation(HHG) to clarify how the temporal coherence of the incident light  is transferred to the high-harmonic radiation field in terms of not only temporal but also spatial coherence.
The study of the coherence transfer through the nonlinear interaction of an atom and/or molecule with the intense laser light may help to understand the  generation of the optical vortex recently experimentally observed.\cite{Gauthier17NatureCom,Taira17NatureCom}

As the interaction with the radiation field increases, the use of the perturbation method becomes inappropriate to describe the dynamics. 
Instead, Floquet method is a powerful tool to investigate the time evolution of a quantum state under a strong periodic external field.\cite{Grifoni98PhysRep,Shirley65PR,Sambe73PRA,Kayanuma94PRA,Mori15PRA}
With use of Floquet method, the time evolution of a system with finite degrees of freedom has been studied so far.
But when we  apply the Floquet method to study the dynamics of open quantum system, such as a spontaneous emission decay process, a serious problem then arises, because we should take into account the interaction with the radiation field with infinite degrees of freedom and we necessarily encounter the resonance singularity.
Actually it is a fundamental problem how to derive the irreversible decay process from time reversible microscopic dynamics.

In order to describe the irreversible process based on the microscopic dynamics, Prigogine and one of the authors (T.P.) {\it et al.} have developed the theory of complex spectral analysis where the function space is extended to {\it the extended Hilbert space}.\cite{Petrosky91Physica,Petrosky96Chaos}
In the extended Hilbert space, the total  {\it Hermitian} Hamiltonian composed of  a subsystem and the environment may have complex eigenvalues representing decay process.
In our preceding works, by using this theory combined with Floquet method, we have investigated the decay dynamics of an impurity embedded in a one-dimensional continuum under a time-periodic external field.\cite{Yamada12PRB,Noba11PRB}

Compared with phenomenological analysis using effective Hamiltonian,\cite{Hatano96PRL,Bender98PRL,Heiss04JPA}  this method has a strong advantage that the correlation between the subsystem and the environment is properly described, since the time evolution of the system is represented by the expansion of the eigenstates of the {\it total } Hamiltonian.
This is essential in the study of the high-harmonic generation, since we have to consider the coherence transfer between the atom and the radiation field in a unified way.
 
In this work, we study the HHG from the excited atom whose energy is sinusoidally changed by a coherent time-dependent external field by using the complex spectral analysis in the extended Floquet-Hilbert space.
We have obtained the complex eigenstates of the total Hamiltonian with use of Feshbach-Brilloiun-Wigner projection method.
Time evolution of the system is represented by the complex eigenvector expansion so that the correlation dynamics between the photon and the atom is fully taken into account. 
We found that the spectral intensity of the HHG is nonlinear to the amplitude of the incident light, as it is determined by the Bessel function in terms of $A/\omega$, where $A$ and $\omega$ are the amplitude and the frequency of the incident field, respectively.
We have also found that the spatial distribution of the emitted radiation shows a pulse-like emission as a result of the interference of the multiple Floquet resonance states.
With use of the complex spectral analysis in the extended Floquet-Hilbert space, it is clearly understood that the temporal coherence of the incident light is transferred to the spatial coherence through the resonance singularity.
Some possible applications to the systems with time-dependent strong interaction between matter and the radiation field are discussed in the conclusion.

\psection{Model and the Floquet Hamiltonian}

We consider the spontaneous emission of an excited state of a two level atom when the excited energy level is sinusoidally  changed by an external field.
In the present work, the Hamiltonian is described by one-dimensional Friedrichs model:  \cite{Yamada12PRB,Friedrichs48}
\begin{align}
\label{eq:H}
H(t)=(\epsilon_{d}+A\sin{\omega t})|d\>\<d|+\sum_{k}\epsilon_{k}|k\>\<k|+\lambda\sum_{k}V_{k}(|d\>\<k|+|k\>\<d|) \;,
\end{align}
where $|d\>$ and $|k\>$ represent the excited state of the atom and an emitted photon state with the energies of $\epsilon_d$ and $\epsilon_k=\hbar c |k|$.
The energy scheme is shown in Figure \ref{fig:energy_scheme}(a).
\begin{figure}[t]
\begin{center}
\includegraphics[width=16cm]{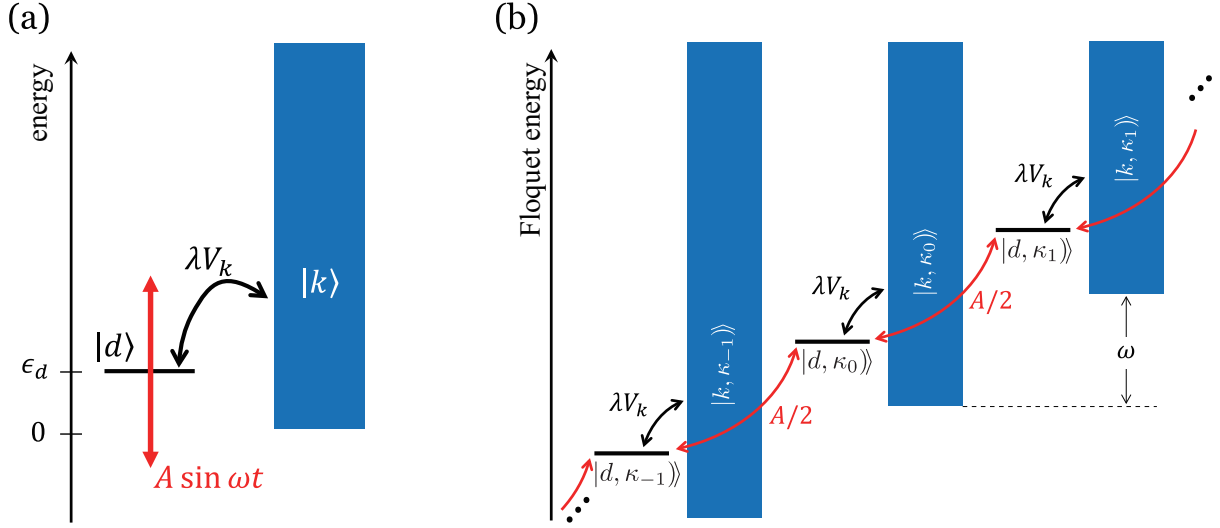}
\caption{(Color online) (a) The unperturbed energy scheme of the $\cal R$ space Hamiltonian Eq.(\ref{eq:H}). 
(b) The unperturbed energy scheme of  the Floquet Hamiltonian Eq.(\ref{eq:FH}).
}
\label{fig:energy_scheme}
\end{center}
\end{figure}
Hereafter we take the units of $\hbar=1$ and $c=1$.
We use the box normalization and take the limit $L\to\infty$, making wave number $k$ continuous variable.
In Eq.(\ref{eq:H}), we use $V_k=(4\pi\epsilon_k/L)^{1/2}\left(1-\theta(\left||k|-|k_c|\right|)\right)$ with a natural cutoff $k_c(=2\pi)$.
We have adopted the rotating-wave-approximation (RWA).

Since the Hamiltonian is periodic in time, $H(t+T)=H(t)$ with a period  of $T\equiv2\pi/\omega$, we  may apply the Floquet theorem to solve the time-dependent Schr{\"o}dinger equation, where the solution of $ i\partial_t|\Psi(t)\>=H(t)|\Psi(t)\>$ is assumed to take the form  of  $|\Psi_{\alpha}(t)\>=e^{-iz t}|\Phi (t)\>$ with a periodic function $|\Phi(t+T)\>=|\Phi (t)\>$.\cite{Grifoni98PhysRep,Shirley65PR,Sambe73PRA}
The substitution of the wave function into the Schr{\"o}dinger equation leads to the  eigenvalue problem of the Floquet Hamiltonian given as
\begin{align}\label{EVPhi}
{\cal H}|\Phi (t)\>\equiv \left[H(t)-i\partial_t \right]|\Phi (t)\>=z |\Phi(t)\> \;.
\end{align}

For the time periodic function $\Phi(t)$, we may introduce {\it the time basis} denoted by $\{|t)\}$ in the ${\cal T}$-space,\cite{Lungu00PhysScr} satisfying the orthonormality and the completeness relations
\begin{align}
(t|t')=T\delta(t-t') \;,\; \hat I_{\cal T}={1\over T}\int_0^T |t)(t| dt \;,
\end{align}
where $\hat I_{\cal T}$ is the unit operator in the ${\cal T}$-space.
The time basis is the eigenstate of the {\it time operator} defined by
\begin{align}
\hat t \equiv {1\over T}\int_0^T t |t)(t| dt \;,
\end{align}
so that $\hat t|t)=t|t)$.
The {\it time wave number operator} is also defined by $\hat p_t\equiv i\partial_t$. 
The eigenstate of $\hat p_t$ is the {\it time wave number basis} given by
\begin{align}
|\kappa_n)\equiv {1\over T}\int_0^T e^{i\kappa_n t}|t) \;, \quad \hat p_t|\kappa_n)=\kappa_n |\kappa_n)\;,
\end{align}
where $\kappa_n=n\omega=2\pi n/T$ $(n=0,1,\cdots)$.
The direct product of the time basis space ($\cal T$-space) and the $\{|d\>, |k\>\}$ atom-radiation-field vector space ($\cal R$-space) forms the  Floquet-Hilbert space, or simply call the Floquet space: ${\cal F}={\cal R}\otimes {\cal T}$ space.\cite{Grifoni98PhysRep,Lungu00PhysScr}

Any state in the Floquet space is represented by the Floquet basis as
\begin{align}
|\Phi)\!\>={1\over T}\int_0^T dt  \sum_{\alpha=d,k} |\alpha,t \)>\(<  \alpha,t|\Phi \)> = \sum_{n=-\infty}^\infty \sum_{\alpha=d,k} |\alpha,\kappa_n )\!\>\<\!(\alpha,\kappa_n|\Phi)\!\> \;.
\end{align}
Note the time periodic state in the Hilbert space (${\cal R}$-space) is given by  $|\Phi (t)\>= |(t|\Phi)\!\>$. 

In terms of the Floquet basis, Eq.(\ref{EVPhi}) is represented as a time-independent eigenvalue problem
\begin{align}\label{FloquetEVP}
{\cal H} |\Phi\)>=z|\Phi\)>  \;,
\end{align}
where the Floquet Hamiltonian is represented by 
\begin{align}
\label{eq:FH}
{\cal H}=&\sum_{n=-\infty}^{+\infty} \left\{ (\epsilon_{d}+\kappa_n) |d,\kappa_n\)> \(<d,\kappa_n|+\sum_{k}(\epsilon_{k}+\kappa_n)|k,\kappa_n\)>\(<k,\kappa_n|  \right.\nonumber\\
&\left. +\lambda \sum_{k} V_k \Bigl[ |k,\kappa_n\)>\(<d,\kappa_n|+|d,\kappa_n\)>\(<k,\kappa_n|  \Bigr]+\frac{A}{2} \Bigl[ |d,\kappa_{n+1}\)>\(<d,\kappa_n|+|d,\kappa_n\)>\(<d,\kappa_{n+1}|  \Bigr] \right\} \;.
\end{align}
The unperturbed energy scheme of Floquet Hamiltonian is shown in Figure \ref{fig:energy_scheme}(b), where there are multiple resonances between the discrete states and the continuum.
This multiple resonance in the Floquet space is essential to cause the pulse-like photon emission as a result of the coherence transfer, as shown below.

\psection{Complex eigenvalue problem of the Floquet Hamiltonian}

In order to describe an irreversible decay from the excited atom emitting a photon, we apply the complex spectral analysis to the present system, where the state vector is represented in the extended Floquet space.\cite{Petrosky91Physica,Petrosky01PRA,Ordonez01PRA}
The right and left eigenvalue problem of $\hat{\cal{H}}$ read
\begin{align}
\label{eq:rightev}
\hat{\cal{H}}|\Phi_\eta^{(m)}\)>=z_\eta^{(m)}|\Phi_\eta^{(m)}\)>,\quad \(<\widetilde{\Phi}_\eta^{(m)}|\hat{\cal{H}}=z_\eta^{(m)}\(<\widetilde{\Phi}_\eta^{(m)}|\;,
\end{align}
where $|\Phi_\eta^{(m)}\)>$ and $\(<\widetilde{\Phi}_\eta^{(m)}|$ are the right and left eigenstates with a common complex eigenvalue $z_\eta^{(m)}$.
It has been proven that $z_{\eta}^{(m)}$ is multiplicative in terms of $\omega$: 
\begin{align}\label{multiplicative}
z_{\eta}^{(m)}=z_\eta+m\omega \;, \quad\text{for} \quad|\Phi_\eta^{(m)}\)>= e^{i \kappa_m \hat t}|\Phi_\eta^{(0)}\)>\;,
\end{align}
where the index $m$ distinguishes the Floquet mode and the mode $m=0$ is especially called {\it the principal mode}.\cite{Grifoni98PhysRep,Lungu00PhysScr}
These eigenstates form a complete biorthonormal set in the extended Floquet space,
\begin{align}
\sum_{m=-\infty}^{+\infty}\sum_{\eta=\tilde d,\tilde k}|\Phi_\eta^{(m)}\)>\(<\widetilde{\Phi}_\eta^{(m)}| =\hat{I},\quad \(<\widetilde\Phi_\eta^{(m)}|\Phi_{\eta'}^{(m')}\)>=\delta^{{\rm Kr}}_{\eta,\eta'}\delta^{{\rm Kr}}_{m,m'}\;,
\end{align}
where indices of $\tilde d$ and $\tilde k$ denote the discrete resonance states and the dressed continuum, respectively.

In order to solve complex eigenvalue problem in the $\cal F$-space, we use the Feshbach-Brilloiun-Wigner projection method by using the projection operators
\begin{align}
P\equiv\sum_{n=-\infty}^{\infty}|d,\kappa_{n}\)>\(<d,\kappa_{n}|,\quad Q\equiv1-P=\sum_{n=-\infty}^{\infty}\sum_{k}|k,\kappa_{n}\)>\(<k,\kappa_{n}|.
\end{align}
Acting these projection operators on Eq.(\ref{eq:rightev}) yields a closed form of the eigenvalue problem of the effective Hamiltonian in the $P$-subspace composed of the $\{ |d,\kappa_n\)>\}$ basis set:
\begin{align}\label{HeffEVP}
{\cal H}_{\rm eff}(z_\eta^{(m)})P|\Phi_\eta^{(m)}\)>=z_\eta^{(m)} P|\Phi_\eta^{(m)}\)> \;,\;  \(< \tilde \Phi_\eta^{(m)}|P{\cal H}_{\rm eff}(z_\eta^{(m)})=z_\eta^{(m)}\<\tilde \Phi_\eta^{(m)}|P  \;,
\end{align}
where the effective Hamiltonian ${\cal H}_{\rm eff}(z)$ is given by 
\begin{subequations}\label{HeffForm}
\begin{align}
{\cal H}_{\rm eff}(z)=&P{\cal H}P+P{\cal H}Q\frac{1}{z-Q{\cal H}Q}Q{\cal H}P\\
=&\begin{pmatrix}
\ddots&&&&\\
&\epsilon_d-\omega+\lambda^2\Sigma^{(-1)}(z)&-A/2i&0&\\
&A/2i&\epsilon_d+\lambda^2\Sigma^{(0)}(z)&-A/2i&\\
&0&A/2i&\epsilon_d+\omega+\lambda^2\Sigma^{(1)}(z)&\\
&&&&\ddots
\end{pmatrix} \;.
\end{align}
\end{subequations}
Note that   the eigenvalue problem is nonlinear in the sense that the effective Hamiltonian depends on its own eigenvalue  in the eigenvalue problem Eq.(\ref{HeffEVP}).
Only when this nonlinearity is taken into account, the eigenvalues of the effective Hamiltonian coincides with those of the total Hamiltonian.\cite{Petrosky91Physica,Tanaka16PRA}
In Eq.(\ref{HeffForm}),  $\Sigma^{(n)}(z)$ is  the self-energy represented by the Cauchy integral in the limit $L\to\infty$:
\begin{align}
\label{eq:selfeng}
\Sigma^{(n)}(z)=\int_{-\infty}^\infty \frac{v_{k}^{2}}{z-n\omega-\epsilon_k} dk \quad  \quad  \left( v_k\equiv\sqrt{L\over 2\pi} V_k \right)\;.
\end{align}
Because of the resonance singularity in the self-energy,  the effective Hamiltonian becomes non-Hermitian possessing the complex eigenvalues.

Even though the effect of the interaction with the radiation field is renormalized into the self energy function, the effective Hamiltonian matrix is still infinite dimensional tridiagonal matrix corresponding to the infinite numbers of Floquet modes.
We have solved the eigenvalue problem of ${\cal H}_{\rm eff}$ by means of the continued fraction method.\cite{Yamada12PRB}

In the week coupling case $\lambda\ll1$, the discrete eigenvalue is analytically obtained by an expansion around $\epsilon_d$ as
\begin{align}
\label{eq:dev}
z_{\tilde d}\equiv z_{\tilde d}^{(0)}\equiv \varepsilon_{\tilde d}+i\gamma_{\tilde d} + O(\lambda^{4})=\epsilon_d+\lambda^{2}\sum_{n=-\infty}^{+\infty}\Sigma^{(n)}(\epsilon_d)J_{n}^{2}\left(A/\omega\right)+O(\lambda^{4})  \;,
\end{align}
where $J_{n}(x)$ is the $n$-th order Bessel function.


Once we have solved the eigenvalue problem of ${\cal H}_{\rm eff}(z)$ in the $P$-subspace, we can obtain the eigenstate of the {\it total} Hamiltonian with the same eigenvalue by adding the $Q$-component:
\begin{align} \label{Phidm}
|\Phi_{\tilde d}^{(m)}\)>=P|\Phi_{\tilde d}^{(m)}\)> + Q|\Phi_{\tilde d}^{(m)}\)>=N_{\tilde d}^{1/2}\sum_{n=-\infty}^\infty R^{(n)}_{\tilde d} \left[|{d},\kappa_n+\kappa_m\)>+\sum_{k}\frac{\lambda V_k}{(z-\epsilon_k)^+_{z_{\tilde d}^{(n)}}}|{k},\kappa_n+\kappa_m\)>\right]\;. 
\end{align}
In Eq.(\ref{Phidm}),   $N_d$ is normalization constant, and  $R^{(n)}_{\tilde d}$ is represented by the continued fraction,whose  explicit forms are given in Refs.\cite{Yamada12PRB,YamadaThesis}. 
Since this Floquet resonance state is obtained as a linear combination of the dressed atom states and the dressed continuous states, the coherence between the atom and the radiation field is properly described.
It should be noted that there appear resonance singularities at the complex poles of  $z_d^{(m)}=z_d+\kappa_m$ corresponding to the multiple resonance states of the Floquet modes in the Cauchy integral in the second term in the limit $L\to \infty$.
In the  Cauchy integral, we need to take the  analytic continuation toward these complex poles  from the upper complex plane .\cite{Petrosky91Physica}

The eigenstates for the dressed radiation field $|\Phi_k^{(m)}\)>$  and $\(<\tilde \Phi_k^{(m)}|$ are also obtained in the same way.
The explicit representation has been given in Ref.\cite{YamadaThesis}.

\psection{High harmonic generation}

With use of the complete set of the eigenstates of the total Hamiltonian in the $\cal F$-space obtained in the preceding section, we may study the space-time evolution of HHG from the exited atom.
Taking into account the orthonormality of the Floquet eigenstates, the wave function  is represented in the $\cal R$-space  by
\begin{align}\label{PsiT}
|\Psi (t)\>=\sum_{\eta=\tilde d,\tilde k}\sum_m e^{-iz_\eta^{(m)} t}|\Phi_\eta^{(m)}(t)\>\<\widetilde{\Phi}_\eta^{(m)}(0)|\Psi (0)\> \;,
\end{align}
where the state vectors in the $\cal R$-space are obtained by taking an inner product of $|\Phi_\eta^{(m)}\)>$ with the time basis  $|t)$: $|\Phi_\eta^{(m)}(t)\>=|(t|\Phi_\eta^{(m)}\)>$.

\begin{figure}[t]
\begin{center}
\includegraphics[width=16cm]{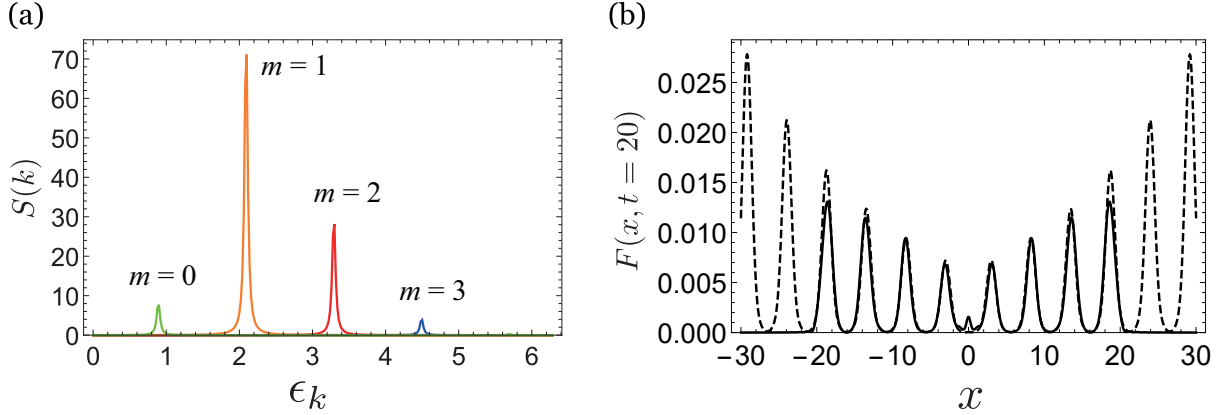}
\caption{(Color online) The HHG spectrum for $\epsilon_d=1.0$, $\omega=1.2$ and $A/\omega=2.0$. (a) Each peak is attributed to the Floquet resonance state: $m=0$ (green), $m=1$ (orange), $m=2$ (red), and $m=3$ (blue).
(b) Spatial distribution of the HHG radiation at $t=20$. The total spectrum is depicted by the solid line, and the resonance state component is depicted by the dashed line. The atom is located at $x=0$.
}
\label{Fig:All}
\end{center}
\end{figure}

The spectrum of the HHG is obtained by 
\begin{align}\label{Skt}
  S(k)=\lim_{t\to\infty}\left|\<k|\Psi(t)\>\right|^2= \left| \sum_{m=-\infty}^\infty  K_d R^{(m)}_{\tilde d} {\lambda V_k \over z_{\tilde d}^{(m)}-\epsilon_k}\right|^2 \simeq  \sum_{m=-\infty}^\infty  \left| K_d R^{(m)}_{\tilde d} \right|^2  {\lambda^2 V_k^2 \over (\varepsilon_{\tilde d}+m\omega-\epsilon_k)^2+\gamma_{\tilde d}^2} \;,\end{align}
where
\begin{align}
K_d\equiv {N_d \over 2\pi} \sum_{n=-\infty}^\infty R_{\tilde d}^{(n)} \;.
\end{align}
As an example, we show in Figure \ref{Fig:All}(a)  the HHG spectrum  for the case with $\epsilon_d=1.0$, $\omega=1.2$ and $A/\omega=2.0$, where 
the spectrum shows several Lorentzian peak profiles at the energy of the  resonance poles $\epsilon_k=\varepsilon_{\tilde d}+m\omega$ and with the width of $\gamma_{\tilde d}$, as given by Eq.(\ref{Skt}).
The intensity of each peak is governed by $ \left| K_d R^{(m)}_{\tilde d} \right|^2\simeq J_{m}^{2}(A/\omega)$,\cite{YamaneResonance} so that the spectral intensity of the HHG is nonlinear to the amplitude of the incident light, as it is determined by the Bessel function in terms of $A/\omega$,  reflecting the renormalized coupling of the dressed atom states with the continuum.\cite{Noba11PRB,YamaneResonance}

The spatial distribution of the HHG is given by 
\begin{subequations}\label{SpaceDist}
\begin{align}
F(x,t)&=|f(x,t)|^2\equiv \left|\<x|\Psi(t)\>\right|^2 \notag\\
&=\left| \sum_m e^{-i z_{\tilde d}^{(m)} t} | \Phi_{\tilde d}^{(m)}(t) \> \<\tilde{\Phi}_{\tilde d}^{(m)}(0) |\Psi (0)\>
 + \sum_{\tilde k}  \sum_m e^{-i z_{\tilde k}^{(m)} t} | \Phi_{\tilde k}^{(m)}(t) \> \<\tilde{\Phi}_{\tilde k}^{(m)}(0) |\Psi (0)\>  \right|^2 \\
&\equiv \left|f_{\tilde d}(x,t)+ \sum_{\tilde k}f_{\tilde k}(x,t) \right|^2  \;,
\end{align}
\end{subequations}
where the first and the second terms  of Eq.(\ref{SpaceDist}b)  are the Floquet resonance states contribution and the Floquet continuous state contribution, respectively. 

As an illustration of the transfer of the temporal coherence of the incident light to the spatial coherence of the emitted field, we show in Figure \ref{Fig:All}(b) by the solid line the spatial distribution of the HHG at $t=20$, $F(x,t=20)$, for the same case with Figure \ref{Fig:All}(a).
The pulse-like photon emission appears as a clear signature of the coherence transfer.

In Figure \ref{Fig:All}(b), we also show the resonance state contribution $|f_{\tilde d}(x,t)|^2$ to the spatial distribution by the dashed line.
We see that inside the light front determined by $x=t$ the total spatial distribution is almost governed by the resonance state contribution, while both differ from each other outside the light front.
It has been revealed that the contribution of the continuous state, $\sum_{\tilde k}f_{\tilde k}(x,t)$,  cancels the spatial distribution outside the light front so that the causality is satisfied.\cite{Petrosky01PRA}

In  order to clarify the origin of the pulse-like photon emission, we shall study the Floquet resonance state contributions $f_{\tilde d}(x,t)$ defined in Eq.(\ref{SpaceDist}b) as
\begin{align}\label{fdxt}
f_{\tilde d}(x,t)\equiv \sum_m e^{-iz_{\tilde d}^{(m)} t}\<x|\Phi_{\tilde d}^{(m)}(t)\>\<\widetilde{\Phi}_{\tilde d}^{(m)}(0)|d\> \;.
\end{align}
With use of Eq.(\ref{Phidm}),   the factor in Eq.(\ref{fdxt})  represented by the $\cal R$ basis is rewritten in terms of the $\cal F$ basis as
\begin{align}\label{xPhidt}
\<x|\Phi_{\tilde d}^{(m)}(t)\>= \(<x,t|\Phi_{\tilde d}^{(m)}\)>=\sum_n \sum_{k}e^{i \kappa_n t}e^{ikx} \(<k,\kappa_n|\Phi_{\tilde d}^{(m)}\)>=\sum_n \sum_{k}e^{i \kappa_n t}e^{ikx} \(<k,\kappa_0|\Phi_{\tilde d}^{(m-n)}\)> \;,
\end{align}
where we have used in the last equality the translation property of Eq.(\ref{multiplicative}).
A simple calculation with use of the Floquet resonance state given by Eq.(\ref{Phidm}) leads to
\begin{align}
\(<k,\kappa_0|\Phi_{\tilde d}^{(m-n)}\)> = N_{\tilde d}^{1/2}R^{(n-m)}_{\tilde d} \frac{\lambda V_k}{(z-\epsilon_k)^+_{z_{\tilde d}^{(n-m)}}}\;. 
\end{align}
Substituting this into Eq.(\ref{fdxt}) with Eq.(\ref{xPhidt}) and replacing the index $n$ by $l=m-n$, we have obtained the resonance state contribution in the limit $L\to \infty$ as
\begin{align}
f_{\tilde d}(x,t)=\lambda \sum_{l=-\infty}^\infty R_{\tilde d}^{(-l)}  e^{-i z_{\tilde d}^{(l)} t} \int_{-\infty}^\infty  {v_k e^{ikx} \over (z-\epsilon_k)^+_{z_{\tilde d}^{(-l)}}} dk  \simeq \lambda \sum_{l=-\infty}^\infty R_{\tilde d}^{(-l)} e^{-i z_{\tilde d}^{(l)} t}  e^{i {z_{\tilde d}^{(-l)}} |x|} \;,
\end{align} 
where we have evaluated the integral by using the residue theorem around the poles of the Floquet resonance states of  $|\Phi_{\tilde d}^{(-l)}\)>$.
The spatial distribution is determined  by the coherent sum of the resonance pole contributions, so that there is a interference between these resonance states in the $\cal F$ space which is given by
\begin{subequations}\label{fint}
\begin{align}
f_{\tilde d}(x,t)^{\rm int} &=\lambda^2 \sum_{l=-\infty}^\infty\sum_{l'=-\infty}^\infty  R_{\tilde d}^{(-l)} [R_{\tilde d}^{(-l)}]^*e^{-i (z_{\tilde d}^{(l)}-z_{\tilde d}^{(l')})(t-|x|)} \\
&=\lambda^2 \sum_{l=-\infty}^\infty\sum_{l'=-\infty}^\infty  R_{\tilde d}^{(-l)} [R_{\tilde d}^{(-l)}]^* e^{\gamma_{\tilde d}(|x|-t)+i(l'-l)\omega(|x|+t)} \;,
\end{align} 
\end{subequations}
where $\gamma_{\tilde d}$ is given by Eq.(\ref{eq:dev}).
The interference term of the Floquet resonance states has a beat between the different Floquet modes accompanied with an exponential growth in space with $\gamma_{\tilde d}$.

We show in Figure \ref{Fig:Reso}(b)  the interference term corresponding to the case with Figure \ref{Fig:All}.
It clearly shows that the effect of this interference is the cause of the pulse-like photon emission in the HHG.
We also show in  Figure \ref{Fig:Reso}(a) the diagonal term of the resonance state contribution which monotonously exponentially grows in space with $\gamma_{\tilde d}$ and the intensity is determined by the $m$-th Bessel function corresponding to the HHG spectrum shown in Figure \ref{Fig:All}(a).

\begin{figure}[t]
\begin{center}
\includegraphics[width=16cm]{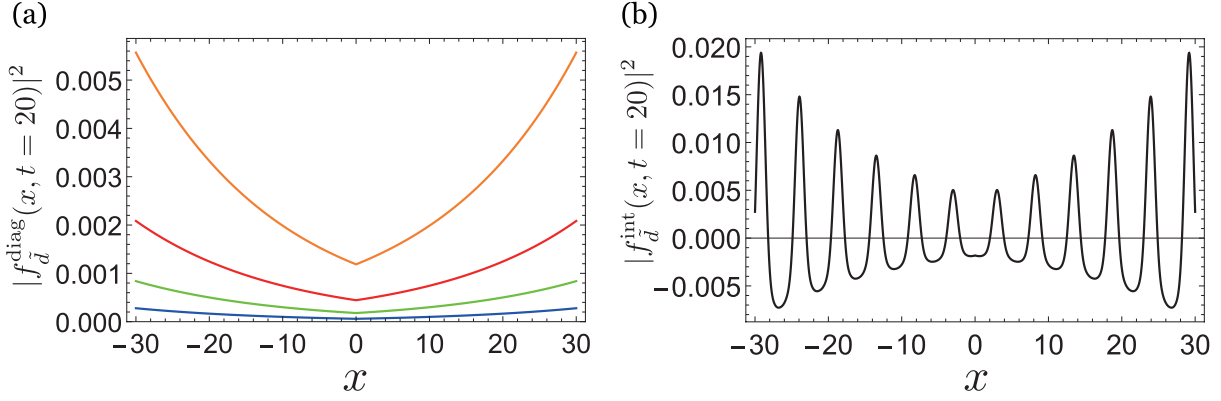}
\caption{(Color online) The spatial distribution of the HHG corresponding to the Floquet resonance states at $t=20$ with the same parameters of Figure \ref{Fig:All}: (a) diagonal term of: $m=0$ (green), $m=1$ (orange), $m=2$ (red), and $m=3$ (blue).
(b) The interference term.}
\label{Fig:Reso}
\end{center}
\end{figure}

\psection{Conclusion}

In this work, we have studied the HHG from the excited atom whose energy is sinusoidally changed by a coherent time-dependent external field by using the complex spectral analysis in the extended Floquet-Hilbert space.
We  obtained the complex eigenstates of the total Hamiltonian with use of Feshbach-Brilloiun-Wigner projection method.
We found that the spectral intensity of the HHG is nonlinear to the amplitude of the incident light, as it is determined by the Bessel function in terms of $A/\omega$.
We have also found that the pulse-like photon emission of the HHG is caused by the interference of the multiple Floquet resonance states.
It shows that the temporal coherence of the incident light is transferred to the spatial coherence through the resonance singularity.

The present analysis can be directly applied to the studies of the time-dependent strong interaction between matter and the radiation field, such as dynamical Casimir effect, \cite{Moore70JMP,Wilson11Nature,Dodonov15JPhys} cavity QED,\cite{Garziano15PRA,Felicetti14PRL}, the generation of the optical vortex, \cite{Gauthier17NatureCom,Taira17NatureCom} and so on.
In order to understand the correlation between the emitted photons and/or between the emitted photons and the material, it is necessary to deal with the material and the radiation field in a unified way.
The present method of the complex spectral analysis in the extended Floquet-Hilbert space is useful tool to investigate these correlations.

\ack

We would like to thank Profs. K. Kanki, K. Noba,  S. Garmon, K. Mizoguchi, and Y. Kayanuma for fruitful discussions.
We would also like to thank Dr. Yamada for providing us with many useful insights.
This work was partially supported by JSPS Grant-in-Aid for Scientific Research Grant Number 16H04003.


%

\end{paper}

\end{document}